\documentclass[journal,twoside,web]{ieeecolor}
\usepackage{generic}
\usepackage{cite}
\usepackage{amsmath,amssymb,amsfonts}
\usepackage{algorithmic}
\usepackage{graphicx}
\usepackage{algorithm,algorithmic}
\usepackage{hyperref}
\hypersetup{hidelinks=true}
\usepackage{textcomp}
\def\BibTeX{{\rm B\kern-.05em{\sc i\kern-.025em b}\kern-.08em
    T\kern-.1667em\lower.7ex\hbox{E}\kern-.125emX}}

\usepackage{multirow} 
\usepackage[caption=false]{subfig}

\usepackage{booktabs} 

\usepackage{graphicx}
\usepackage{overpic}

\begin{document}
%
\title{Female-RHINO: A Real-Time Scanner-Integrated Framework for Automated Quantitative Uterine MRI Analysis and Structured Reporting}

\author{Deepak Bhatia,
Saad Ahmad,
Smiti Tripathy,
Maria Camila Bustos Vivas,
Lieselotte Kratzsch,
Anika Knupfer,
Jordina Aviles Verdera,
Susanne Schulz-Heise,
Matthias May,
and Jana Hutter
\thanks{D Bhatia, S Ahmad, S Tripathy, M Bustos Vivas, L Kratzsch, A Knupfer, J Aviles Verdera, S Schulz-Heise, M May, and J Hutter are with Radiologisches Institut, Universitätsklinikum Erlangen, Erlangen, Germany.}%
\thanks{J Aviles Verdera and J Hutter are with School of Biomedical Engineering and Imaging, KCL, London, UK}
\thanks{J Aviles Verderas, A Knupfer and J Hutter are with Institute for Information Processing, Leibniz University Hannover, Germany and CAIMed, L3S research institute, Hannover, Germany}
\thanks{M May is with Imaging Science Institute, Uniklinik Erlangen, Ulmenweg 18, 91054 Erlangen}
}

\maketitle


\begin{abstract}
Standardized assessment of uterine MRI remains challenging due to anatomical variability, observer dependence, and the lack of workflow-integrated automated analysis tools. This work presents Female-RHINO: (R)eproductive (H)ealth (I)maging A(N)alysis T(O)ol, a real-time AI-assisted framework for automated quantitative uterine MRI analysis and structured reporting during image acquisition. We present an end-to-end system that integrates inline communication with the MRI scanner and deep learning–based analysis to derive quantitative uterine biomarkers from sagittal T2-weighted pelvic MRI. The framework combines segmentation and anatomical landmark detection models trained and evaluated on $>500$ multi-center datasets spanning diverse protocols, vendors, and patient populations. It performs volumetry, detects and quantifies common incidental findings such as fibroids and Nabothian cysts, and extracts six anatomical landmarks for biometric assessment. Results are compiled into a structured, clinician-oriented report with integrated visualizations, without manual interaction. Evaluation on independent retrospective and prospective cohorts demonstrated robust performance across varying acquisition settings. Mean Dice similarity coefficients were 0.82 for the uterus, and 0.80 for fibroids, with lower but consistent agreement for Nabothian cysts. Landmark detection achieved a mean radial error of 3.7 mm. End-to-end processing was completed in under 70 seconds, enabling availability of results during the ongoing scan. Prospective deployment yielded immediate, standardized, and reproducible analyses supported by inter-observer agreement. The proposed system enables real-time, scanner-integrated AI for automated uterine MRI analysis and reporting, with potential to improve standardization, efficiency, and clinical workflow in pelvic imaging.
\end{abstract}


\section{Introduction}
\label{sec:introduction}

Uterine MRI plays an important complementary role to ultrasound (US) in the diagnosis, characterization, and treatment monitoring of a broad spectrum of gynecological conditions, including infertility-related assessment, malignant diseases such as endometrial and cervical cancer, and benign disorders such as fibroids, adenomyosis, endometriosis, and Nabothian cysts~\cite{Agostinho2017-ut,Omi2024-je,patel2010imaging,salman2024magnetic,Raffone2024MRILeiomyomasSarcomas,theis2023deep}. Quantitative assessment of uterine morphology and lesions is clinically relevant for diagnosis, treatment planning, and follow-up of these conditions~\cite{berczi2025outlier,kurban2021uterine,theis2023deep,sevindik2023differences}. Adenomyosis is commonly associated with focal or diffuse thickening of the endometrium and the junctional zone~\cite{Agostinho2017-ut,Zand2007-yf}. Uterine fibroids, also referred to as myomas, are common benign lesions affecting over 30\% of women and are frequently associated with symptoms such as abnormal menstrual bleeding and pelvic pain~\cite{khan2014uterine,stewart2017epidemiology}. Depending on their size and location, fibroids may additionally contribute to pelvic pressure, urinary symptoms, venous congestion, and uterine enlargement~\cite{khan2014uterine,stewart2017epidemiology}. Nabothian cysts are another common incidental finding; although benign and typically asymptomatic, their differentiation from other cervical lesions remains clinically important~\cite{Omi2024-je}. In this context, uterine and lesion volumetry, shape descriptors, corpus length, fundal thickness, anteroposterior diameter, utero-cervical angle, and uterine position provide clinically meaningful descriptors for standardized assessment, treatment planning, and follow-up in gynecological imaging~\cite{berczi2025outlier,kurban2021uterine,sevindik2023differences,xholli2022angle,franconeri2018structured,fidan2017value}.

MRI provides large-volume pelvic coverage, excellent soft-tissue contrast, and - compared to US - reduced operator dependence~\cite{Agostinho2017-ut,pan2024large,salman2024magnetic}. These properties make it ideally suited for visualizing uterine anatomy, characterizing lesions, and performing quantitative measurements. However, uterine MRI assessment remains challenging due to substantial anatomical variability, physiological changes over the menstrual cycle, tissue-air interfaces, bowel-related artifacts, and dynamic motion such as peristalsis~\cite{Kido2008-zv,sevindik2023differences,tong2020recommendations,Zand2007-yf,pan2024large}. In routine practice, quantitative assessment is often performed manually after completion of the examination, which is time-consuming and subject to inter-observer variability~\cite{berczi2025outlier,sevindik2023differences}. Moreover, unlike US, conventional MRI workflows do not provide interactive quantitative feedback during the ongoing examination. These limitations are further amplified by narrative reporting, which can reduce standardization, comparability across examinations, and reproducibility of follow-up assessment~\cite{franconeri2018structured}.

Automated deep learning-based methods have recently shown promise for improving the reproducibility and objectivity of uterine MRI analysis, particularly for segmentation, volumetry, landmark detection and biometric measurements~\cite{Koike2025-oo,kurata2019automatic,Khaghani2024-yh,Khaghani2024-pa,liu2025deep,theis2023deep,mulliez2023three}. However, most existing approaches remain offline, focus on individual tasks such as segmentation or landmark detection, and do not provide an integrated quantitative assessment of uterine anatomy, lesions, and biometric measurements within the clinical acquisition workflow. This limits their potential for real-time clinical use, where immediate quantitative information could support acquisition planning, same-session review, and standardized reporting.

Real-time integration of AI analysis during MRI acquisition has the potential to enable workflow-aware and adaptive imaging strategies, such as triggering additional sequences or optimizing imaging planes based on initial findings, as demonstrated in cardiac, fetal, and pelvic MRI applications~\cite{chow2021prototyping,silva2025automatic,verdera2025heron,xue2019gadgetron,Koike2025-oo}. For uterine MRI, immediate access to quantitative information such as uterine volume, lesion burden, biometric measurements, and characterization of benign findings may support responsive adjustments during the scan, reduce the need for follow-up imaging, and facilitate standardized reporting~\cite{anneveldt2021lessons,chow2021prototyping,liu2025artificial,franconeri2018structured}.

In this study, we present Female-RHINO, a novel real-time scanner-integrated framework for automated quantitative uterine MRI analysis and structured reporting during image acquisition. The proposed system combines deep learning-based 3D segmentation and anatomical landmark detection to automatically derive uterine volumetry, lesion characterization, lesion counts, and biometric measurements from sagittal T2-weighted pelvic MRI. By integrating directly into the MRI acquisition workflow through inline scanner communication, the framework generates clinician-oriented quantitative reports with visualizations without manual interaction. The proposed approach was trained and evaluated on more than 500 multi-center MRI examinations spanning heterogeneous acquisition protocols, field strengths, vendors, and patient populations, and was prospectively deployed for real-time evaluation during pelvic MRI examinations. A preliminary version of this work was presented at the MICCAI 2025 Workshop on Computer-Aided Pelvic Imaging for Female Health~\cite{10.1007/978-3-032-05825-6_13}.

\begin{figure*}[t]
\centering
\begin{overpic}[width=0.4\textwidth]{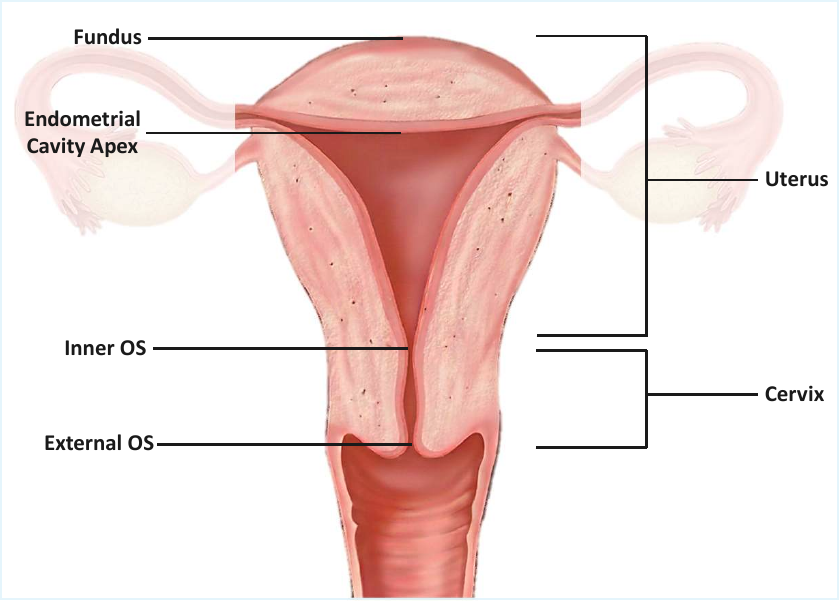}
  \put(0,73){\textbf{(A)}}
\end{overpic}
\hspace{0.6\fill}
\begin{overpic}[width=0.325\textwidth]{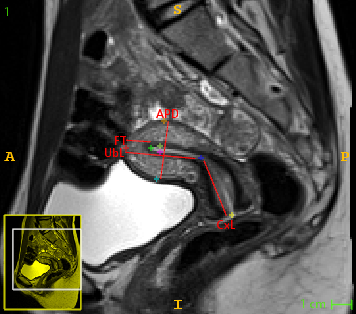}
  \put(1,93){\textbf{(B)}}
\end{overpic}

\vspace{0.15cm}

\begin{overpic}[width=0.8\textwidth]{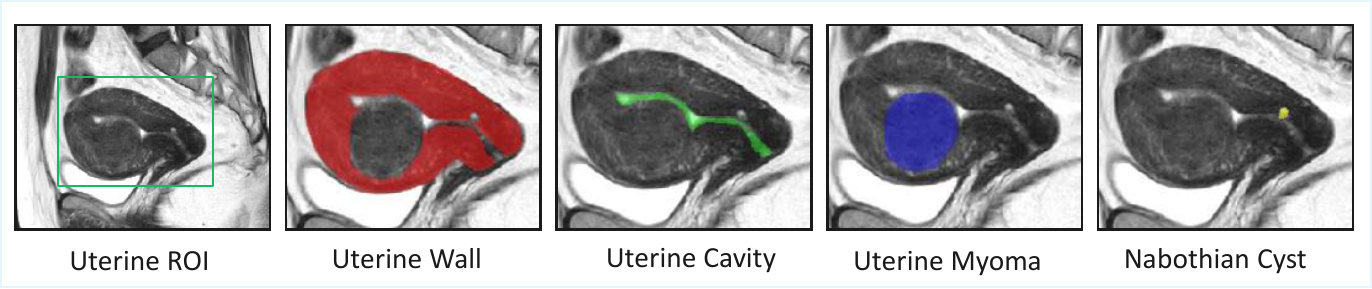}
  \put(2,21){\textbf{(C)}}
\end{overpic}

\caption{Overview of the anatomical and structural information used in this work:
(A) schematic illustration of uterine anatomy, major morphological sections, and key landmarks;
(B) landmarks employed for automated biometry computation on a sagittal T2-weighted MRI, with crosses indicating landmark locations and red lines denoting derived biometric measurements, FT: fundal thickness (Fundus-Cavity Apex), UbL: uterine body length (Fundus-Inner Os), CxL: cervical length (Inner Os-External Os), APD: anterior--posterior diameter;
(C) pixel-level annotations of the four semantic classes used for training and evaluation, including uterine wall (red), uterine/endometrial cavity (green), myomas (blue), and Nabothian cysts (yellow), modified from~\cite{pan2024large}.}
\label{fig:labels}
\end{figure*} 

\section{Methods}

\begin{figure*}[!t]
    \centering
    \includegraphics[width=0.8\textwidth]{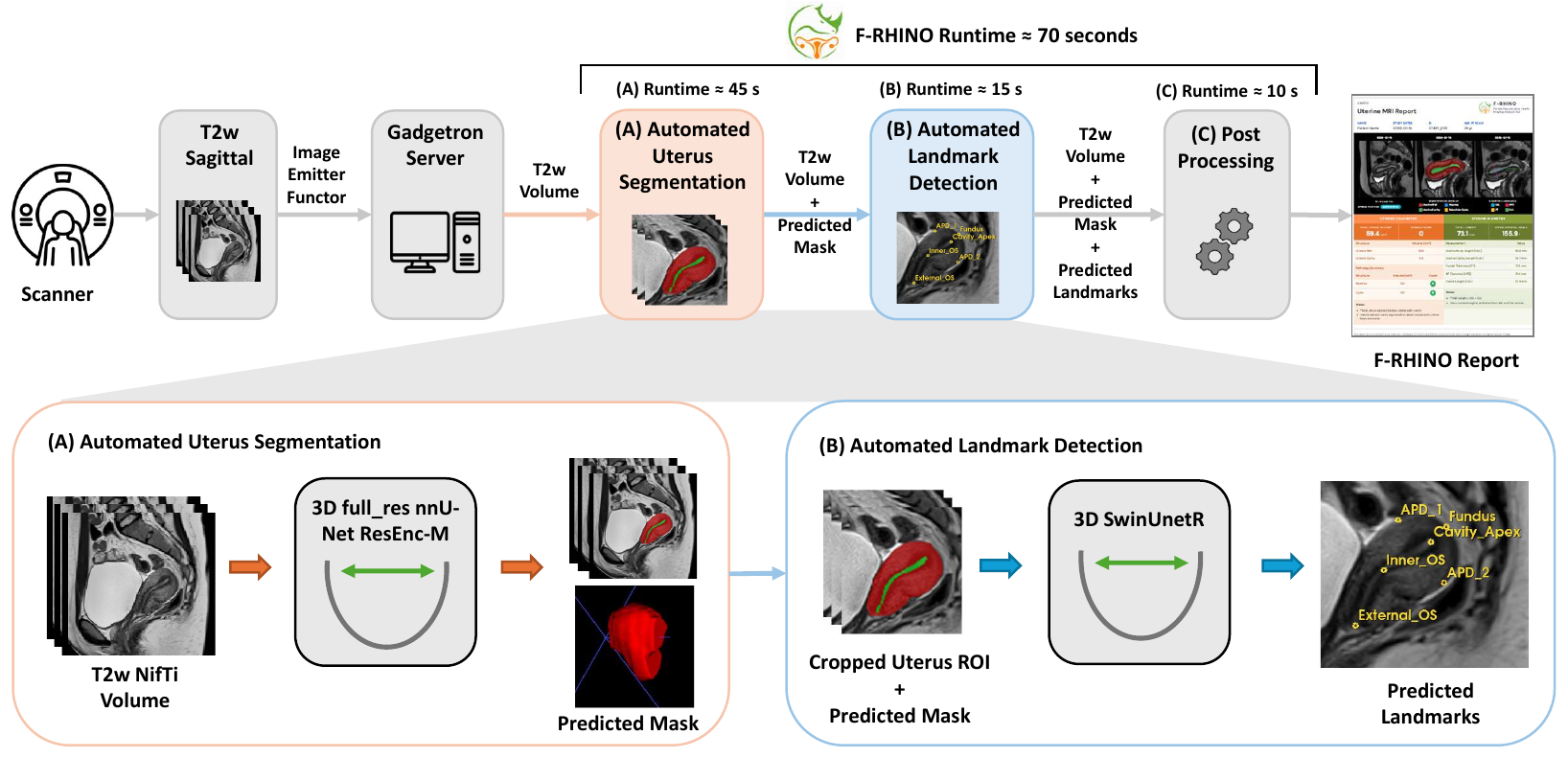}
    \caption{Schematic overview of the workflow implemented in the proposed F-RHINO tool, illustrating the process from data acquisition to generation of the structured quantitative uterine MRI analysis report in the scanner room.}
    \label{methods_pipeline}
\end{figure*}

\noindent In the following, the workflow of the proposed tool from data acquisition to real-time structured report generation is detailed, with a graphical overview given in Figure~\ref{methods_pipeline}.

\begin{table}[t]
\centering
\caption{Cohort demographics and acquisition parameters for the three T2-weighted MRI protocols. Values are reported as mean $\pm$ standard deviation or range where applicable.~\cite{pan2024large}}
\label{tab:mri_params}
\resizebox{\columnwidth}{!}{
\begin{tabular}{lccc}
\toprule
\textbf{Parameter} & \textbf{Protocol I} & \textbf{Protocol II} & \textbf{Protocol III} \\
\midrule
\# Subjects & 300  & 95 & 126 \\
Age (years) & $49.3 \pm 13.0$ & $26.5 \pm 9.4$ & $27.2\pm 5.2$ \\
BMI (kg/m$^2$) & $-$ & $23.0 \pm 4.7$ & $23.5 \pm 3.8$ \\
Field strength (T) & 1.5 -- 3.0 & 0.55 -- 3 & 0.55 \\
TR (ms) & 820 -- 5200 & 1400 / 5800 & 850 / 5700 \\
TE (ms) & 80 -- 130 & 100 / 115 & 110 / 125 \\
Slice thickness (mm) & 4.0 -- 7.5 & 3.0 -- 4.5 & 3.0 -- 4.5 \\
Inter-slice gap (mm) & 4.4 -- 7.5 & 0 & 0 \\
In-plane spacing (mm) & 0.25 -- 0.75 & 0.5 -- 0.98 & 0.5 -- 0.94 \\
FOV (mm$^2$) & $240 \times 240$ & $250 \times 250$ & $240 \times 240$ \\
Scanner / Vendor & Philips & Siemens & Siemens \\
\bottomrule
\end{tabular}
}
\end{table}


\subsection{Data}
\noindent Sagittal T2-weighted MRI is the most commonly used clinical sequence for assessing uterine morphology~\cite{sevindik2023differences,Khaghani2024-pa}. Data from three protocols comprising sagittal T2-weighted pelvic MRI volumes acquired across multiple clinical scanners from different vendors, different magnetic field strengths and acquisition parameters, was used to train and evaluate the proposed tool. Acquisition parameters for all three Protocols, including magnetic field strength, repetition time (TR), echo time (TE), voxel spacing, field of view (FOV) and slice thickness, are summarized in Table~\ref{tab:mri_params}.  

Protocol I consists of a large publicly available dataset of T2-weighted sagittal female pelvic MRI volumes acquired on a Philips 3.0T scanner from 300 patients diagnosed with uterine myomas, of whom 127 also present with Nabothian cysts~\cite{pan2024large}. The dataset covers all nine fibroid subtypes defined by the International Federation of Gynecology and Obstetrics (FIGO) classification~\cite{fraser2011figo,pan2024large}.

Protocol II includes retrospectively collected T2-weighted sagittal MRI volumes (comprised of HASTE and TSE sequences) from female pelvic MRI scans performed across multiple sites and acquisition workflows, with magnetic field strengths ranging from 0.55T to 3T. The cohort consists of an approximately equal proportion of healthy controls and patients with uterine myomas or Nabothian cysts. Only examinations with complete uterine coverage and without substantial motion or reconstruction artifacts were retained. 

Protocol III comprises a prospective dataset consisting T2-weighted HASTE and TSE sequences, acquired as part of an ongoing research study after informed consent was obtained (23-444-Bm) at 0.55T using a Siemens MAGNETOM Free.Max scanner (Siemens Healthineers, Forchheim, Germany). This cohort reflects the typical contrast characteristics of low-field MRI and was therefore used to evaluate the robustness of the proposed automated analysis tool under low SNR conditions.


\begin{figure*}[!t]
    \centering
    \includegraphics[width=0.8\textwidth]{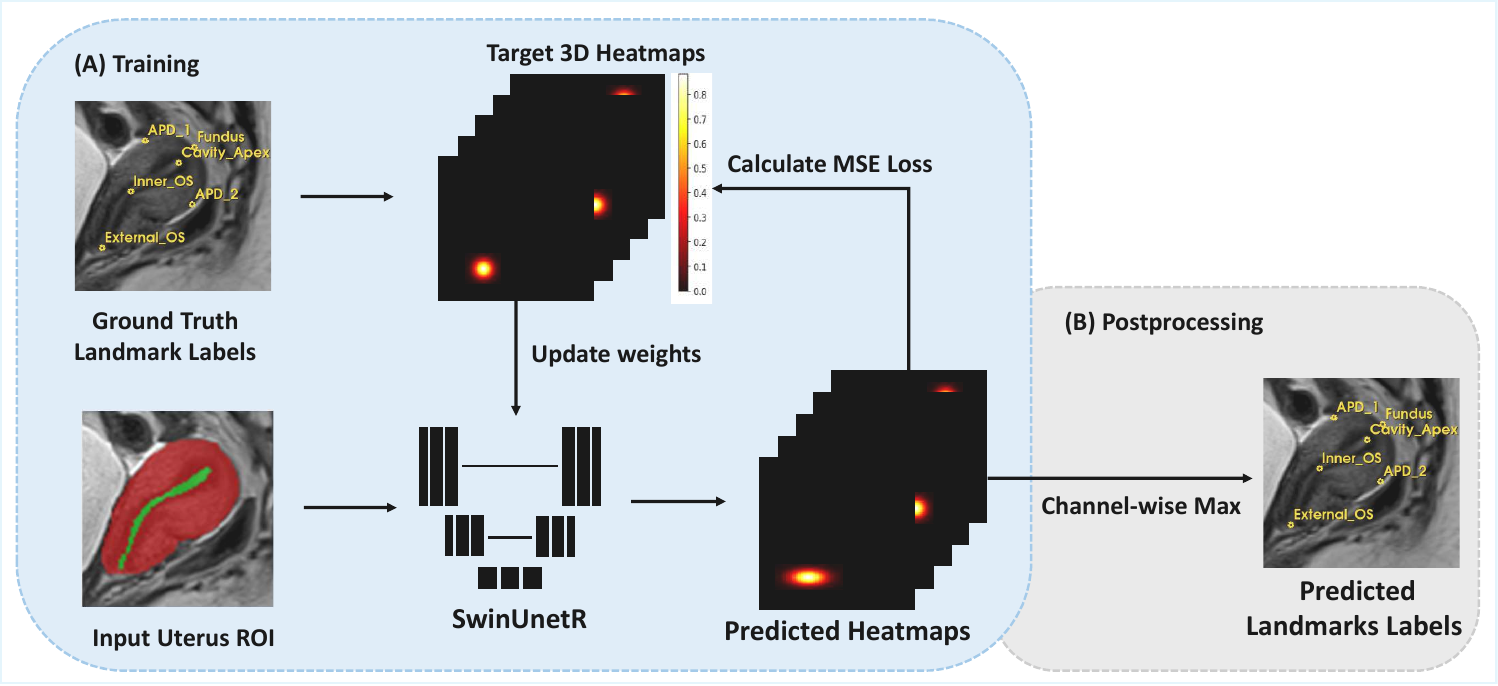}
\caption{Depiction of the steps followed in the landmark training. From left to right: Acquired data with ground truth labels and uterine ROI masks, transformed into target 3D heatmaps and training using the SwinUnetR to iteratively obtain predicted heatmaps. Finally in (B), predicted landmarks are obtained by applying channel-wise max on the predicted heatmaps.}
\label{landmark_training}
\end{figure*}
Each protocol includes segmentation labels for four clinically relevant structures: the uterine wall, the uterine cavity, uterine myomas, and Nabothian cysts. These segmentation labels are illustrated in Fig.~\ref{fig:labels}C. All segmentation masks were manually annotated using ITK-SNAP~\cite{py06nimg} following a standardized annotation protocol developed jointly by expert gynecologic radiologists and pelvic MRI experts. To ensure consistency across protocols, anatomically ambiguous cases were reviewed and resolved by expert consensus. 

In addition to segmentation labels, a set of anatomically defined landmarks for automated uterine biometry was annotated. As shown in Fig.~\ref{fig:labels}A--B, these landmarks include the fundus, endometrial cavity apex, internal os, external os, and bilateral anteroposterior diameter points, defined following established gynecological imaging conventions to ensure consistent anatomical interpretation across subjects and datasets~\cite{sevindik2023differences}. For landmark annotation, anatomically constrained geometric priors were used to derive reproducible landmark candidates from the uterine segmentation masks. The uterine cavity was represented by a skeletonized centerline capturing the anatomical course of the endometrial and cervical canal. Candidate terminal points of this centerline, together with the segmentation masks, were used to identify the fundus, cavity apex, and external os, while distance-transform-based spatial cues were used to distinguish the cavity apex from the cervical endpoint. The internal os was estimated along the cavity centerline by identifying the anatomically narrowest transition region between the uterine cavity and cervical canal. For anteroposterior diameter estimation, the principal uterine axis defined by the fundus and internal os landmarks served as an anatomical reference, and the APD endpoints were selected as the longest wall-to-wall chord perpendicular to this axis within the uterine wall mask. All automatically generated landmark candidates were reviewed by a clinical expert, with anatomically inconsistent cases corrected manually and additional manual annotations performed when reliable automatic estimates were not available. This strategy enabled standardized and reproducible landmark annotation while preserving anatomical plausibility across heterogeneous uterine shapes and acquisition protocols, although cases with large fibroids in Protocol I occasionally required increased manual correction due to substantial distortion of normal uterine anatomy.

\subsection{Automated Uterine Segmentation and Volumetry}\label{seg_vol}
\noindent Automatic segmentation was performed using nnUNet-v2 with a 3D full resolution Residual Encoder UNet (ResEnc-UNet-M) configuration~\cite{isensee2024nnu}. The self-configuring nnUNet framework was used, which automatically determines preprocessing, network architecture, training hyperparameters, and data augmentation based on the training data characteristics. The input MRI volumes were normalized using z-score normalization and resampled to a median voxel spacing of $4.4 \times 0.48 \times 0.48\,\mathrm{mm}$. Training was performed using patches of size $16 \times 320 \times 256$ voxels and a batch size of 2. The network was trained using a combined dice and cross-entropy loss and optimized using stochastic gradient descent with Nesterov momentum and an initial learning rate of $10^{-2}$. Polynomial learning rate decay was applied with a weight decay of $3\times10^{-5}$ and deep supervision was enabled at multiple decoder levels. Five-fold cross-validation was employed for training and during inference, predictions from all folds were ensembled by averaging the softmax probability maps from each fold. All models were trained from scratch for 1000 epochs on a single NVIDIA A40 GPU. In total, 392 MRI volumes were used for training and validation, and 51 volumes for testing, across the three protocols. We derived volumetric measurements from the predicted segmentation masks which are reported in milliliters (\(\text{mL}\)). 3D connected component analysis was applied to the predicted masks of the uterine myoma and Nabothian cyst classes. Each in this way identified spatially contiguous cluster of labeled voxels was treated as a distinct lesion to enable computation of lesion-wise statistics.

\subsection{Automated Uterine Landmark Detection and Biometry}\label{biometry}
\noindent Landmark detection was performed using a Swin-UNETR~\cite{hatamizadeh2021swin} network implemented using PyTorch~\cite{paszke2019pytorch}, trained to localize uterine anatomical landmarks within a cropped region of interest (ROI) encompassing the uterus. All input MRI volumes were first resampled to an isotropic resolution of $1.0\,\mathrm{mm}$. Following resampling, the uterus ROI was extracted with a $10\,\mathrm{mm}$ margin around the uterus segmentation mask in all spatial directions. The resulting volumes were normalized to zero mean and unit standard deviation. We applied random scale of intensity in the range (0.9, 1.1) to the input ROI volumes. During training, each landmark ground truth label was converted into a 3D Gaussian blob of range $(0, 1)$ and a standard deviation of $\sigma = 4$. The number of network output channels was set to five, with one channel jointly representing the two APD landmarks in order to avoid the model to confuse between the two diameter points and one channel each for the remaining four anatomical landmarks. The model was trained using a Mean Squared Error (MSE) loss to regress the continuous heatmap values. Training was carried out for 300 epochs using the Adam optimizer with an initial learning rate of $1 \times 10^{-3}$ and a \texttt{ReduceLROnPlateau} scheduler, with a batch size of 1 on an NVIDIA A100 GPU. At inference, the trained network produced continuous 3D heatmaps for each landmark channel. Landmark coordinates were defined by the peaks of channel-wise maxima. An overview of the automated landmark detection process is shown in Figure~\ref{landmark_training}. In total, 215 MRI volumes were used for training, 25 for validation, and 45 for testing across the three datasets. From the predicted landmark coordinates, uterine body length (UbL), cervical length (CxL), anteroposterior diameter (APD), and fundal thickness (FT) were computed in millimetres. The utero-cervical angle and uterine flexion were derived from the geometric relationship between the UbL and CxL vectors. Based on their relative orientation, the uterus position was classified as anteverted or  retroverted~\cite{fidan2017value}.

\subsection{Online Deployment and Automatic Report Generation}
\noindent The connection between acquisition and analysis is obtained via  the Fire interface~\cite{chow2021prototyping,xue2019gadgetron}, enabling real-time execution of the proposed tool at the time of the MR acquisition. The tool was implemented on a 0.55T MAGNETOM Free.Max scanner (Siemens Healthineers, Forchheim, Germany), equipped with the FIRE tool and the Gadgetron framework connected to a dedicated workstation equipped with an 11GB NVIDIA GEFORCE RTX 2080 Ti GPU. During MRI acquisition, T2-weighted sagittal DICOM slices are streamed in real time from the scanner computer to the GPU server in ISMRMD format. The slices are stacked together and converted into a NifTi file for further processing. The trained segmentation model then performs automated segmentation on the NifTi file, generating segmentation masks and post-processing steps compute volumetric measurements for each class and identify and enumerate individual lesions as detailed in section~\ref{seg_vol}~\cite{isensee2024nnu}. In addition, anatomical landmarks are automatically localized using the trained landmark detection model on the Uterus ROI cropped through the predicted segmentation masks and uterine biometry as detailed in section~\ref{biometry} is derived. Finally, a structured HTML report as shown in Figure~\ref{report}, was automatically generated using Python. The report was designed as a clinician-oriented summary of the automated analysis, integrating representative sagittal images, segmentation overlays, landmark-based biometry, uterine and lesion volumetry, pathology counts, and contextual reference-range information in a standardized layout~\cite{franconeri2018structured}. The visual overlays allow review of the predicted segmentation, landmark locations, and derived measurements. Reference ranges for uterine volume and biometric measurements were derived from published normative data and are provided for contextual comparison rather than as diagnostic thresholds~\cite{kelsey2016validated,sevindik2023differences}. The age-related uterine volumetry reference data presented in~\cite{kelsey2016validated} is in line with recent large-population normative data reported for subjects in their twenties to forties~\cite{Khaghani2024-pa,Khaghani2024-yh}. 
\begin{table*}[t]
\centering
\caption{Segmentation performance (mean $\pm$ SD) on test sets from Protocol~I, Protocol~II, and Protocol~III, reported using Dice Similarity Coefficient (DSC) and Intersection over Union (IoU). The number of evaluated cases per structure is reported as $n$.}
\label{tab:results}

\begin{tabular*}{0.85\textwidth}{@{\extracolsep{\fill}}lccccccccc}
\toprule
& \multicolumn{3}{c}{\textbf{Protocol I (n=30)}} 
& \multicolumn{3}{c}{\textbf{Protocol II (n=11)}} 
& \multicolumn{3}{c}{\textbf{Protocol III (n=10)}} \\
\cmidrule(lr){2-4} \cmidrule(lr){5-7} \cmidrule(lr){8-10}

\textbf{Structure} 
& \textbf{DSC$\pm$Std} & \textbf{IoU$\pm$Std} & \textbf{$n$}
& \textbf{DSC$\pm$Std} & \textbf{IoU$\pm$Std} & \textbf{$n$}
& \textbf{DSC$\pm$Std} & \textbf{IoU$\pm$Std} & \textbf{$n$} \\
\midrule

Uterine Wall    
& 0.83 $\pm$ 0.11 & 0.71 $\pm$ 0.12 & 30
& 0.82 $\pm$ 0.10 & 0.71 $\pm$ 0.14 & 11
& 0.86 $\pm$ 0.05 & 0.75 $\pm$ 0.07 & 10 \\

Uterine Cavity  
& 0.79 $\pm$ 0.12 & 0.66 $\pm$ 0.13 & 30
& 0.79 $\pm$ 0.11 & 0.66 $\pm$ 0.14 & 11
& 0.80 $\pm$ 0.07 & 0.68 $\pm$ 0.10 & 10 \\

Uterine Myoma         
& 0.76 $\pm$ 0.27 & 0.66 $\pm$ 0.26 & 30
& 0.83 $\pm$ 0.12 & 0.72 $\pm$ 0.17 & 7
& 0.81 $\pm$ 0.13 & 0.69 $\pm$ 0.19 & 2 \\

Nabothian Cyst 
& 0.52 $\pm$ 0.38 & 0.44 $\pm$ 0.36 & 14
& 0.30 $\pm$ 0.29 & 0.20 $\pm$ 0.21 & 4
& 0.61 $\pm$ 0.27 & 0.48 $\pm$ 0.29 & 4 \\
\bottomrule
\end{tabular*}
\end{table*}
\begin{figure*}[!t]
    \centering
    \includegraphics[width=0.75\textwidth]{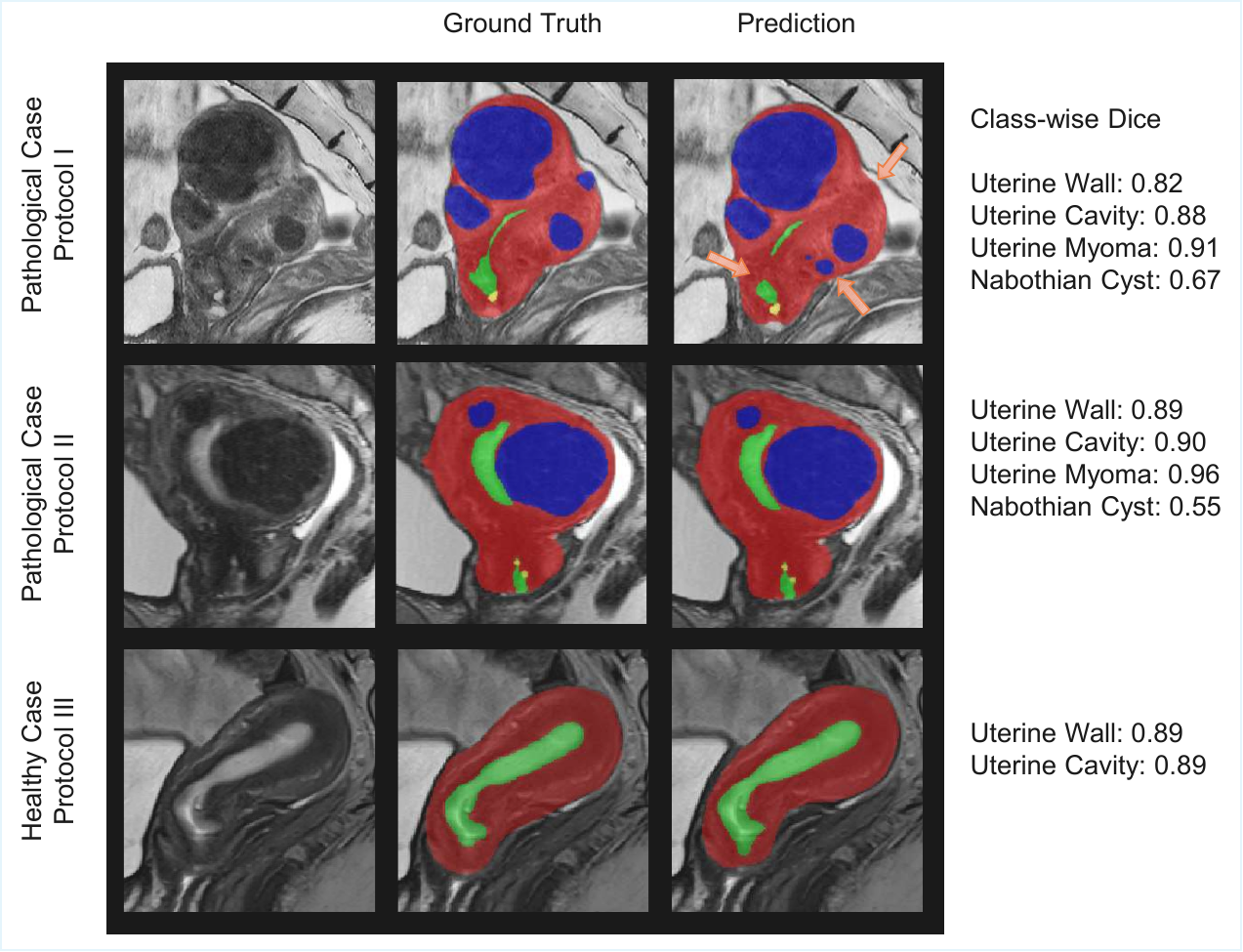}
\caption{Qualitative comparison of segmentation results, showing ground truth and predicted masks for a pathological case from Protocol I (top row), a pathological case from Protocol II (middle row) and a healthy case from Protocol II (bottom row). The ground truth masks and predicted masks highlight anatomical structures including the uterine wall (red), uterine cavity (green), uterine myoma (blue), and Nabothian cyst (yellow)}
\label{segmentation_figure}
\end{figure*}

\subsection{Evaluation}
\subsubsection{Automated Segmentation and landmark detection}
\noindent The evaluation of the automated segmentation was performed on the basis of segmentation accuracy using two widely adopted metrics: the Dice Similarity Coefficient (DSC) and the Intersection over Union (IoU). The DSC quantifies the volumetric overlap between the predicted and reference masks, whereas the IoU measures the ratio of the intersecting to the union regions. Both metrics were computed per-segmentation class for the held-out test set comprised of 51 subjects from the three protocols, and subsequently averaged within each anatomical label. To assess landmark detection for anatomical plausibility, we applied a hard radial-error threshold of 10 mm. Predictions exceeding this threshold were considered failed detections. This threshold was chosen to separate clear localization failures from smaller deviations while accounting for variability in slice thickness across the included MRI protocols. Results are reported across 45 test subjects from the three protocols. Using this failure criterion, we calculated the failure count for each landmark and protocol. For all remaining valid detections, mean radial error, standard deviation, and median radial error were computed to quantify localization accuracy.

\subsubsection{Prospective Analysis}
\noindent The proposed tool was prospectively evaluated in real time on five cases acquired with Protocol~III. For each case, all automated uterine volumetric and biometric measurements were compared against two independently performed manual measurements. Quantitative agreement was assessed using comparisons between both manual measurements (inter-observer variability), and manual and automatic measurements. For volumetric analysis, absolute volume differences and Dice similarity coefficients were computed between corresponding segmentation masks and for biometric measurements, the accuracy was quantified using the mean absolute difference. Finally, the complete and component wise inference time for each case, from image acquisition to automated report generation was recorded.

\begin{table*}[t]
\centering
\footnotesize
\caption{Landmark localization performance across the three datasets.
Mean radial error (MRE) $\pm$ standard deviation and median radial error (MdRE) are computed
on non-failure cases. Failures are defined as localization error $>10$\,mm and are
reported as a fraction of the test set size per protocol.}
\label{tab:landmark_protocol}

\begin{tabular*}{0.8\textwidth}{@{\extracolsep{\fill}}lccc ccc ccc}
\toprule
& \multicolumn{3}{c}{\textbf{Protocol I}}
& \multicolumn{3}{c}{\textbf{Protocol II}}
& \multicolumn{3}{c}{\textbf{Protocol III}} \\
\cmidrule(lr){2-4} \cmidrule(lr){5-7} \cmidrule(lr){8-10}

\textbf{Landmark}
& \textbf{MRE$\pm$Std} & \textbf{MdRE} & \textbf{Fail}
& \textbf{MRE$\pm$Std} & \textbf{MdRE} & \textbf{Fail}
& \textbf{MRE$\pm$Std} & \textbf{MdRE} & \textbf{Fail} \\
& [mm] & [mm] &
& [mm] & [mm] &
& [mm] & [mm] & \\
\midrule

Cavity Apex
& 4.22$\pm$2.52 & 4.75 & 0/15
& \textbf{2.71$\pm$1.55} & \textbf{2.42} & 0/15
& 2.69$\pm$1.44 & 2.42 & 0/15 \\

Inner Os
& 4.03$\pm$2.30 & \textbf{3.13} & 0/15
& 4.13$\pm$1.94 & 3.63 & 1/15
& \textbf{3.77$\pm$2.25} & 3.40 & 0/15 \\

External Os
& \textbf{2.75$\pm$1.42} & 2.59 & 2/15
& 2.88$\pm$1.31 & \textbf{2.35} & 1/15
& 3.61$\pm$1.73 & 3.40 & 0/15 \\

Fundus
& \textbf{3.59$\pm$2.64} & \textbf{2.84} & 1/15
& 3.98$\pm$2.45 & 3.22 & 0/15
& 3.67$\pm$2.56 & 3.24 & 0/15 \\

APD-1
& 4.78$\pm$2.71 & 3.39 & 1/15
& \textbf{3.69$\pm$2.21} & 3.96 & 0/15
& 4.22$\pm$1.83 & \textbf{3.91} & 0/15 \\

APD-2
& 4.28$\pm$2.23 & \textbf{3.40} & 0/15
& \textbf{3.79$\pm$1.93} & 3.73 & 0/15
& 3.89$\pm$2.20 & 3.44 & 0/15 \\

\midrule
\textbf{Overall}
& 3.96$\pm$2.37 & 3.18 & 4/90
& \textbf{3.53$\pm$1.96} & \textbf{3.16} & 2/90
& 3.64$\pm$2.03 & 3.40 & \textbf{0/90} \\
\bottomrule
\end{tabular*}
\end{table*}

\section{Results}
\subsection{Automated Segmentation}
\noindent Figure~\ref{segmentation_figure} illustrates a qualitative comparison between ground truth masks and predicted masks for both healthy and pathological cases across the three datasets. As shown in Table~\ref{tab:results},  segmentation performance of the uterine wall achieved the highest accuracy across all datasets, with mean DSC values of 0.83, 0.82, and 0.86 for datasets~I, II, and III, respectively. Similar trends were observed for the uterine cavity, with consistent performance across datasets. Segmentation of uterine myomas exhibited increased variability, reflected by larger standard deviations, particularly in Protocol~I. Nabothian cyst segmentation showed the lowest overall accuracy and largest variability across all datasets.
\begin{figure*}[!t]
    \centering
    \includegraphics[width=0.75\textwidth]{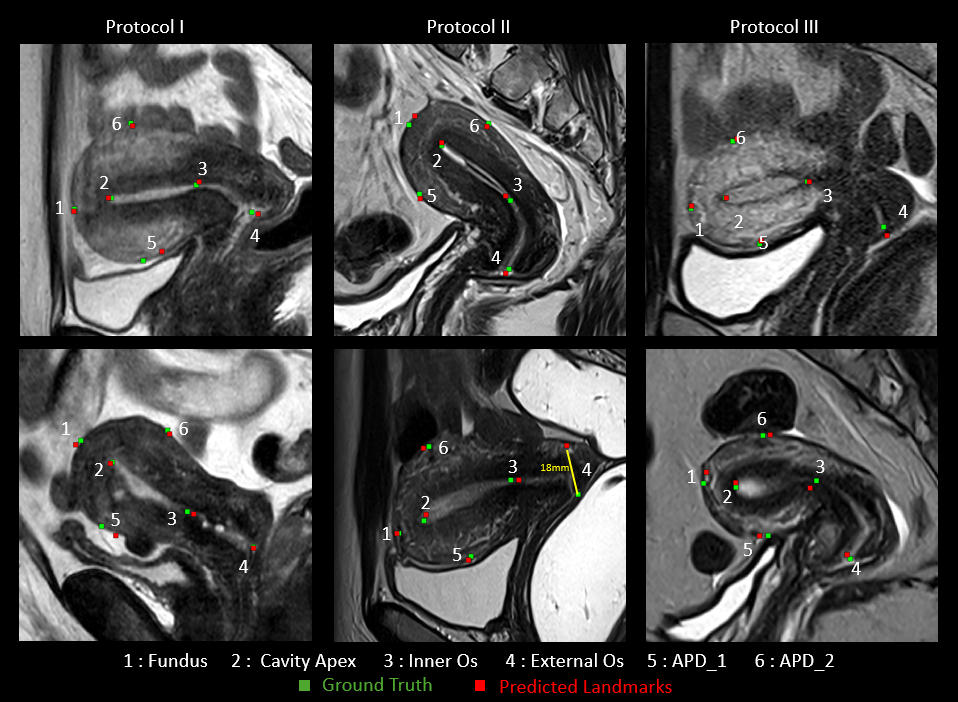}
\caption{Qualitative comparison of landmark detection results, showing ground truth and predicted landmarks}
\label{landmark_preds}
\end{figure*}
\subsection{Automated Landmark Detection}
\noindent Figure~\ref{landmark_preds} shows a qualitative comparison between ground truth landmarks and predicted landmarks for 45 cases encompassing all three datasets. As shown in Table IV, detection on Protocol I achieved an overall MRE of 3.96$\pm$2.37 mm with an MdRE of 3.18 mm and 4 failures out of 90 evaluated landmarks. Protocol II yielded an overall MRE of 3.53$\pm$1.96 mm with an MdRE of 3.26 mm and 2 failures out of 90 landmarks. Protocol III showed MRE of 3.64$\pm$2.03 mm and an MdRE of 3.40 mm, with 0 failures out of 90 landmarks. Aggregating results across all datasets, the proposed method achieved a pooled MRE of 3.71 $\pm$2.12 mm and a pooled MdRE of 3.21 mm, with 6 failures out of 270 landmarks.
\begin{table}[t]
\centering
\caption{Dice Similarity Score (DSC) and Absolute Volume Difference (AVD) in ml between manual and automated uterine volumetric measurements in the prospective evaluation ($n=5$). Inter-observer variability is reported between manual observers (R1 \& R2), alongside comparisons with the automated method.}
\label{tab:prospective_volume}
\begin{tabular}{lccc}
\toprule
\textbf{Comparison} & \textbf{DSC$\pm$Std} & \textbf{AVD$\pm$Std} \\
& & [ml]\\
\midrule
R1--R2 & 0.83 $\pm$ 0.06 & 5.3 $\pm$ 5.2 \\
R1--Automatic & 0.85 $\pm$ 0.03 & 5.8 $\pm$ 1.6 \\
R2--Automatic & 0.85 $\pm$ 0.07 & 7.3 $\pm$ 7.3 \\
\bottomrule
\end{tabular}
\end{table}

\begin{table}[t]
\centering
\caption{Mean absolute differences (in mm) between manual and automated uterine biometric measurements in the prospective evaluation ($n=5$). Inter-observer variability is reported between manual observers (R1 \& R2), alongside comparisons with the automated method.}
\label{tab:prospective_biometry}
\begin{tabular}{lccc}
\toprule
\textbf{Measurement} 
& \textbf{R1--R2} 
& \textbf{R1--Auto} 
& \textbf{R2--Auto} \\
&[mm] & [mm] & [mm] \\
\midrule
Fundal thickness (FT)                 & 1.18$\pm$1.36 & 1.58$\pm$1.77&\textbf{0.85$\pm$0.46} \\
Uterine body length (UbL)             & 3.98$\pm$2.54 &  \textbf{2.06$\pm$1.62} & 3.47$\pm$2.94 \\
Cervical length (CxL)                 & \textbf{1.85$\pm$1.02} & 3.56$\pm$1.97 & 3.20$\pm$2.58 \\
Anteroposterior diameter (APD)        & \textbf{1.02$\pm$1.55} & 2.12$\pm$1.47& 1.28$\pm$0.92 \\
\midrule
\textbf{Average} 
& \textbf{2.01$\pm$1.17} 
&  \textbf{2.33$\pm$1.61}
&  \textbf{2.20$\pm$1.05} \\
\bottomrule
\end{tabular}
\end{table}

\begin{figure*}[t]
    \centering
    \includegraphics[width=0.4\textwidth]{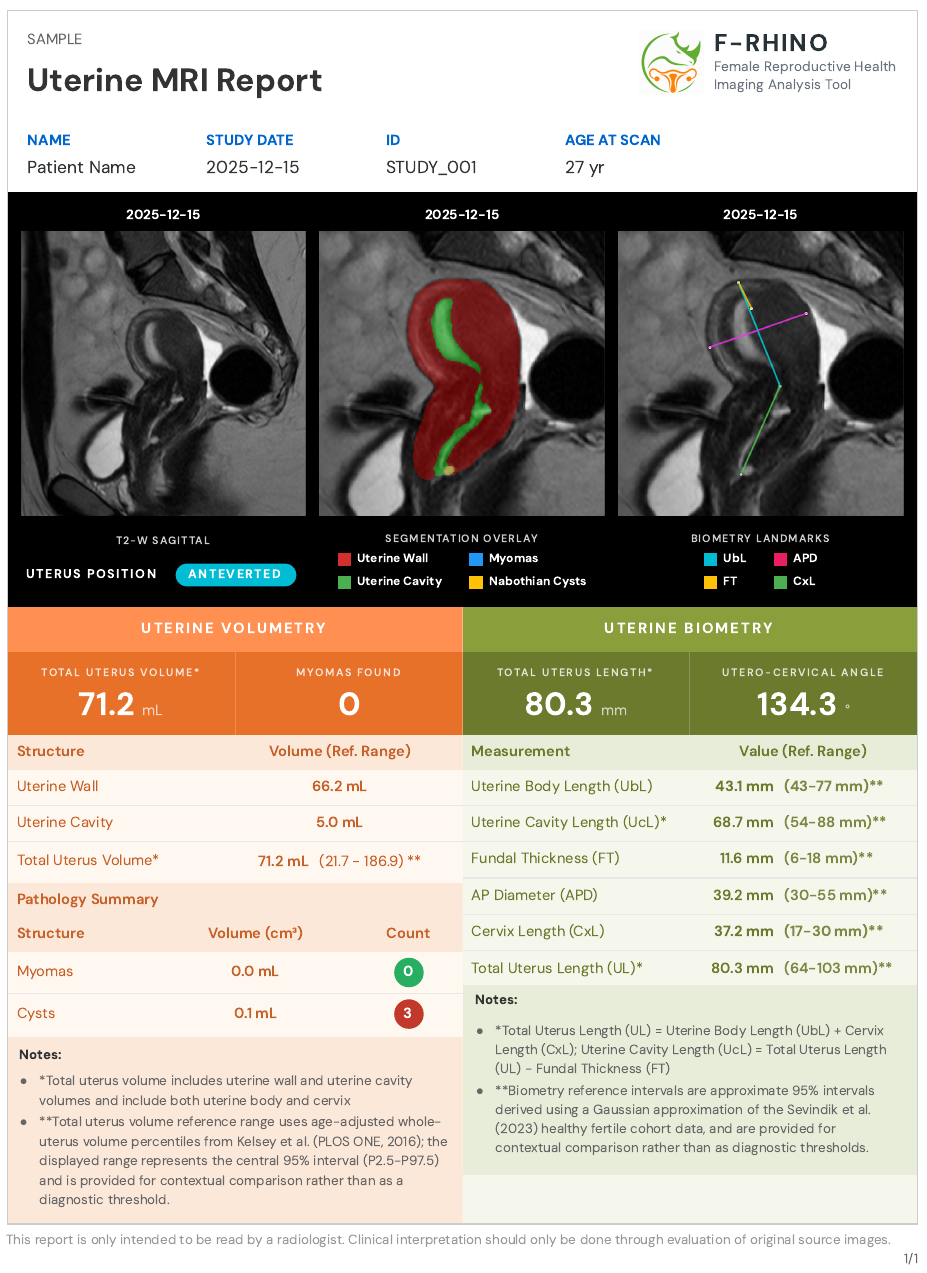}
    \includegraphics[width=0.4\textwidth]{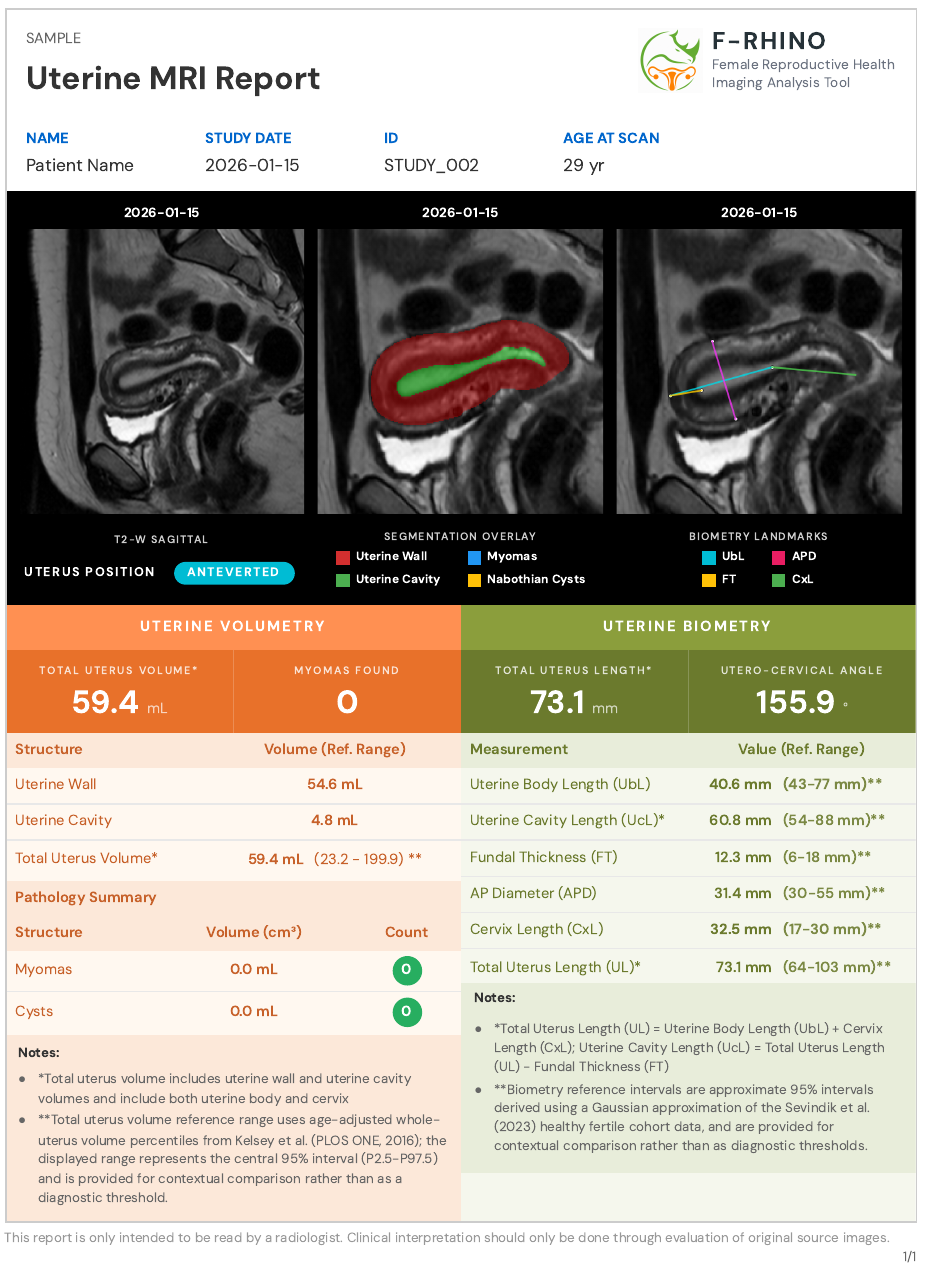}
    \caption{Prospective examples of automated uterine MRI analysis reports generated via F-RHINO. The reports present automated segmentation-based uterine volumetry, automated landmark-derived uterine biometry and position, visualizations on representative slices, and characterization of incidental findings such as myomas and Nabothian cysts, if detected. The report is fully generated in under 70 seconds using the proposed tool.}
    \label{report}
\end{figure*}

\subsection{Prospective Testing}
\noindent Figure~\ref{report} illustrates prospectively generated F-RHINO uterine MRI analysis reports in real time. Prospective evaluation results are summarized in Tables~\ref{tab:prospective_biometry} and~\ref{tab:prospective_volume}. For volumetric analysis, automated segmentations showed strong agreement with both manual references. Dice similarity coefficients for automatic versus manual comparisons were comparable to inter-observer agreement (0.85$\pm$0.03 and 0.85$\pm$0.07 vs. 0.83$\pm$0.06). Absolute volume differences followed a similar trend, with automatic-to-manual differences of 5.8$\pm$1.6\,ml and 7.3$\pm$7.3\,ml compared with an inter-observer difference of 5.3$\pm$5.2\,ml, indicating that automated volumetry performed consistently with manual annotations. For biometric measurements, absolute differences between automated and manual measurements were 2.33$\pm$1.61\,mm and 2.20$\pm$1.05\,mm, respectively, compared with an inter-observer difference of 2.01$\pm$1.17\,mm. No landmark detection failures were observed in the prospective testing. Measurement-wise, fundal thickness and anteroposterior diameter showed the closest agreement with manual measurements, whereas uterine body length and cervical length showed slightly larger differences, likely related to the shared dependence of both measurements on internal os localization, which can be challenging on MRI~\cite{liu2024study,Koike2025-oo}. End-to-end inference time remained stable across all evaluated cases, with automated report generation completed within $70$\,s. The component-wise run time break down has been provided in Figure~\ref{methods_pipeline}.


\section{Discussion}

 Our proposed framework enables real-time, scanner-integrated uterine MRI analysis with structured report generation during image acquisition. By automatically deriving uterine volumetry, lesion burden, and landmark-based biometric measurements within a single integrated analysis, the system moves beyond task-specific offline methods and provides clinically relevant quantitative information in a structured report during the examination. This addresses a key limitation of routine uterine MRI assessment, where uterine and lesion volume estimation often relies on simplified geometric approximations derived from manual biometric measurements or manual slice-by-slice contouring. Such measurements are time-consuming, observer-dependent, and commonly documented in narrative reports, which may limit reproducibility and longitudinal comparability, particularly in cases with multiple lesions or distorted anatomy~\cite{kelsey2016validated,sevindik2023differences,theis2023deep,franconeri2018structured,berczi2025outlier}. The retrospective and prospective evaluations show feasiblity to deploy the proposed framework within clinical uterine MRI workflows in the future.

The reported DSC and IoU scores are in line with literature addressing similar multiclass uterine MRI segmentation problems~\cite{kurata2019automatic,Khaghani2024-yh,Khaghani2024-pa,Pan2023WAResUNet}, supporting the use of the proposed tool for automated uterine volumetry~\cite{Khaghani2024-pa}. Uterine wall, cavity, and myoma segmentation showed robust performance across heterogeneous datasets, whereas Nabothian cyst segmentation remained more challenging, consistent with previously reported trends~\cite{Pan2023WAResUNet}. These results primarily reflect inherent differences between the segmented structures, ranging from large myomas with minimal surrounding healthy tissue to very small myomas, as well as small or absent Nabothian cysts~\cite{pan2023instance,Pan2023WAResUNet,pan2024large}. Such variability can lead to partial or missed predictions in challenging cases, particularly in Protocol~I, thereby disproportionately reducing the average DSC~\cite{maier2024metrics}. Compared with previous studies largely focused on single datasets or controlled acquisition settings~\cite{kurata2019automatic,pan2023instance,Saleem2025DeepSegmentationUterineMyomas,Pan2023WAResUNet,pan2024large,Khaghani2024-pa,theis2023deep}, the present study evaluates performance in a heterogeneous multi-center setting spanning multiple acquisition protocols, field strengths, and MRI sequences. Protocol-dependent differences were therefore expected, with Protocol~II and III, which predominantly contained healthy uteri and mild benign findings with clearer anatomical boundaries, showing improved myoma segmentation performance compared with Protocol~I. Importantly, the prospective evaluation further demonstrated that automated uterine volumetry was comparable to inter-observer variability, with automatic-to-manual DSC values of 0.85 and absolute volume differences of 5.8--7.3\,ml, compared with an inter-observer DSC of 0.83 and an absolute volume difference of 5.3\,ml~\cite{maier2024metrics}.

Directly comparable work on automated multi-landmark detection for uterine MRI remains limited~\cite{Koike2025-oo,mulliez2023three}. Nevertheless, the landmark detection results demonstrated consistent localization accuracy across heterogeneous MRI datasets and were in line with heatmap-regression-based 3D landmark localization approaches in pelvic and medical MRI~\cite{Koike2025-oo,payer2019integrating,pang2024prior}. In clinical practice, uterine biometric measurements are commonly performed on selected sagittal images and therefore depend on manual slice selection and landmark placement~\cite{sevindik2023differences}. By localizing anatomical landmarks in 3D physical space, the proposed approach reduces dependence on a single image slice and accounts for out-of-plane anatomical variation, particularly in rotated uteri or cases with pathology-related anatomical distortion. The low overall failure rate, defined as MRE $>10$\,mm, indicates robust landmark localization despite variability in image quality, acquisition parameters, and anatomical appearance. Error differences across protocols were partly attributable to acquisition characteristics. In Protocol~I, distorted anatomy caused by fibroids limited the number of cases in which biometry could be reliably measured for training and evaluation. In addition, the larger slice thickness and inter-slice gap reduced through-plane resolution, increasing uncertainty for small or boundary-dependent landmarks. This interpretation is supported by recent work on automated uterine landmark detection for pelvic MRI prescription, where coordinate-wise landmark deviations showed higher through-plane variability~\cite{Koike2025-oo}. In contrast, Protocols~II and III showed lower localization errors and fewer failures, potentially reflecting their lower slice thickness and absence of inter-slice gaps. Landmark-specific variations mainly reflected anatomical ambiguity and sensitivity to partial-volume effects, particularly for the fundus, cervical, and anteroposterior diameter landmarks. Importantly, the prospective evaluation showed that automated biometric measurements were comparable to inter-observer variability, with average automatic-to-manual differences of 2.33$\pm$1.61\,mm and 2.20$\pm$1.05\,mm compared with an inter-observer difference of 2.01$\pm$1.17\,mm. Overall, these results indicate that the proposed framework enables reliable uterine landmark localization across acquisition protocols and supports automated extraction of clinically relevant biometric measurements.

A key strength of the proposed framework is its real-time integration of automated analysis outputs into a structured, clinician-oriented report during MRI acquisition. Rather than providing isolated segmentation masks or landmark coordinates, the system translates volumetry, lesion characterization, lesion counts, uterine biometry, and representative visualizations into standardized quantitative documentation. Compared with conventional narrative reporting, this may improve reproducibility and facilitate longitudinal comparison across examinations~\cite{franconeri2018structured}. The integrated visualizations further allow radiologists to verify the automated measurements, supporting transparency and clinical plausibility. Another strength is the use of heterogeneous data from multiple clinical sites, scanners, acquisition protocols, and magnetic field strengths, which introduced realistic domain shifts during training and evaluation. Despite this variability, the framework maintained stable segmentation and landmark detection performance across protocols, supporting its robustness and generalizability. Although implemented as a scanner-integrated system for real-time use, the framework can also be used offline for retrospective analysis or conventional post-acquisition evaluation. While the generated report is not intended to replace comprehensive radiological interpretation, it provides automated quantitative support for clinical reporting and acquisition planning~\cite{franconeri2018structured,silva2025automatic}. The current focus on uterine myomas and Nabothian cysts reflects their high prevalence and the clinical need to differentiate benign findings from malignant lesions~\cite{Takahama2025DifferentialDiagnosisCervicalCysticLesions,Raffone2024MRILeiomyomasSarcomas}. Nevertheless, the framework is flexible and may be extended toward automated detection, volumetry, and characterization of additional benign and malignant uterine pathologies.

This study, however, also has several limitations. The segmentation labels did not distinguish uterine layers such as the endometrium, junctional zone, and myometrium, and the uterine wall and cavity labels included both the uterine body and cervix, preventing cervix-specific volumetry~\cite{Agostinho2017-ut,Khaghani2024-pa}. The current implementation is restricted to sagittal T2-weighted MRI, and the prospective evaluation was limited to a small feasibility cohort. Cases with substantial pathology-related anatomical distortion, particularly large fibroids, remained more challenging for landmark localization and biometry. Future work will focus on layered uterine segmentation, cervix-specific analysis, improved landmark robustness in anatomically challenging cases, and adaptive acquisition strategies aligned to uterine orientation and pathology distribution.

\section{Conclusion}
\noindent We present a fully automated scanner-integrated framework for quantitative uterine MRI analysis that combines deep learning-based 3D segmentation, anatomical landmark detection, and real-time structured report generation during image acquisition. The proposed framework enables rapid extraction of clinically relevant uterine measurements, including volumetry, lesion characterization, lesion counts, and biometric parameters, without manual interaction. By automating a time-intensive and observer-dependent workflow, the system has the potential to improve efficiency, standardization, and reproducibility in routine gynecological MRI assessment. Prospective deployment demonstrated the feasibility of real-time AI-assisted uterine MRI analysis within the clinical acquisition workflow, with automated quantitative reports generated during the ongoing scan. The proposed framework represents a step toward workflow-aware and adaptive pelvic MRI, where quantitative AI analysis can directly support image acquisition, clinical interpretation and personalized diagnostic decision-making in clinical settings, particularly in the evaluation of uterine anomalies, preoperative planning and reproductive health.

\section*{Disclosure of Interests}
\noindent The authors have no competing interests to declare that are relevant to the content of this article.

\section*{Acknowledgments}
\noindent The authors thank all women for participating in this study. This work was supported by the High Tech Agenda Bavaria, DFG Heisenberg funding [502024488], the ERC Starting grant EARTHWORM [101165242] and funding from the Bavarian Minstry for Health, Prevention and Care [EndoKI project]. Parts of this study were funded by NUM 2.0 (FKZ: 01KX2121) and NUM 3.0 (FKZ: 01KX2524).

\section*{References}


\bibliographystyle{IEEEtran}
\bibliography{bibtex/bib/deepak}

@article{maier2024metrics,
  title={Metrics reloaded: recommendations for image analysis validation},
  author={Maier-Hein, Lena and Reinke, Annika and Godau, Patrick and Tizabi, Minu D and Buettner, Florian and Christodoulou, Evangelia and Glocker, Ben and Isensee, Fabian and Kleesiek, Jens and Kozubek, Michal and others},
  journal={Nature methods},
  volume={21},
  number={2},
  pages={195--212},
  year={2024},
  publisher={Nature Publishing Group US New York}
}

@article{fidan2017value,
  title={Value of vaginal cervical position in estimating uterine anatomy},
  author={Fidan, Ula{\c{s}} and Keskin, U{\u{g}}ur and Ulubay, Mustafa and {\"O}zt{\"u}rk, Mustafa and Bodur, Serkan},
  journal={Clinical Anatomy},
  volume={30},
  number={3},
  pages={404--408},
  year={2017},
  publisher={Wiley Online Library}
}

@INPROCEEDINGS{Koike2025-oo,
  title           = "Robust {3D} landmark detection framework for one-stop
                     automated pelvic {MRI} prescription",
  booktitle       = "{ISMRM} Annual Meeting",
  author          = "Koike, Tatsuki and Kudo, Akira and Fuchigami, Takuya and
                     Tachibana, Atsushi and Ikegawa, Ayaka and Yokohama, Wataru
                     and Sakuragi, Kenta and Kitamura, Yoshiro and Hori,
                     Masatoshi and Tomiyama, Noriyuki",
  publisher       = "ISMRM",
  number          =  1395,
  year            =  2025,
  address         = "Concord, CA",
  conference      = "2025 ISMRM \& ISMRT Annual Meeting",
  location        = "Honolulu, Hawaii, USA"
}

@InProceedings{10.1007/978-3-032-05825-6_13,
author="Bhatia, Deepak
and Aviles Verdera, Jordina
and Kitzberger, Michael
and Tripathy, Smiti
and Bustos Vivas, Maria Camila
and Kratzsch, Lieselotte
and Knupfer, Anika
and Hutter, Jana",
title="Real-Time Automated Analysis and Reporting of Uterine MRI",
booktitle="Skin Image Analysis, and Computer-Aided Pelvic Imaging for Female Health",
year="2026",
publisher="Springer Nature Switzerland",
pages="137--147"
}

@article{kurata2019automatic,
  title={Automatic segmentation of the uterus on MRI using a convolutional neural network},
  author={Kurata, Yasuhisa and Nishio, Mizuho and Kido, Aki and Fujimoto, Koji and Yakami, Masahiro and Isoda, Hiroyoshi and Togashi, Kaori},
  journal={Computers in biology and medicine},
  volume={114},
  pages={103438},
  year={2019},
  publisher={Elsevier}
}

@article{liu2024study,
  title={Study on the value of MRI in locating the internal OS of the cervix and influencing factors},
  author={Liu, Mingming and Liang, Yuting and Zheng, Xingzheng and Mo, Na and Jin, Erhu},
  journal={Scientific Reports},
  volume={14},
  number={1},
  pages={17784},
  year={2024},
  publisher={Nature Publishing Group UK London}
}

@article{payer2019integrating,
  title={Integrating spatial configuration into heatmap regression based CNNs for landmark localization},
  author={Payer, Christian and {\v{S}}tern, Darko and Bischof, Horst and Urschler, Martin},
  journal={Medical image analysis},
  volume={54},
  pages={207--219},
  year={2019},
  publisher={Elsevier}
}

@inproceedings{pang2024prior,
  title={Prior guided 3D medical image landmark localization},
  author={Pang, Yijie and Cheng, Pujin and Lyu, Junyan and Lin, Fan and Tang, Xiaoying},
  booktitle={Medical Imaging with Deep Learning},
  pages={1163--1175},
  year={2024},
  organization={PMLR}
}

@article{pan2023instance,
  title={An instance segmentation model based on deep learning for intelligent diagnosis of uterine myomas in mri},
  author={Pan, Haixia and Zhang, Meng and Bai, Wenpei and Li, Bin and Wang, Hongqiang and Geng, Haotian and Zhao, Xiaoran and Zhang, Dongdong and Li, Yanan and Chen, Minghuang},
  journal={Diagnostics},
  volume={13},
  number={9},
  pages={1525},
  year={2023},
  publisher={MDPI}
}

@article{kelsey2016validated,
  title={A validated normative model for human uterine volume from birth to age 40 years},
  author={Kelsey, Thomas W and Ginbey, Eleanor and Chowdhury, Moti M and Bath, Louise E and Anderson, Richard A and Wallace, W Hamish B},
  journal={PloS one},
  volume={11},
  number={6},
  pages={e0157375},
  year={2016},
  publisher={Public Library of Science San Francisco, CA USA}
}

@article{xholli2022angle,
  title={Angle of uterine flexion and adenomyosis},
  author={Xholli, Anjeza and Scovazzi, Umberto and Londero, Ambrogio Pietro and Evangelisti, Giulio and Cavalli, Elena and Schiaffino, Maria Giulia and Vacca, Ilaria and Oppedisano, Francesca and Ferraro, Mattia Francesco and Sirito, Giorgio and others},
  journal={Journal of clinical medicine},
  volume={11},
  number={11},
  pages={3214},
  year={2022},
  publisher={MDPI}
}

@article{franconeri2018structured,
  title={Structured vs narrative reporting of pelvic MRI for fibroids: clarity and impact on treatment planning},
  author={Franconeri, Andrea and Fang, Jieming and Carney, Benjamin and Justaniah, Almamoon and Miller, Laura and Hur, Hye-Chun and King, Louise P and Alammari, Roa and Faintuch, Salomao and Mortele, Koenraad J and others},
  journal={European radiology},
  volume={28},
  number={7},
  pages={3009--3017},
  year={2018},
  publisher={Springer}
}

@article{mulliez2023three,
  title={Three-dimensional measurement of the uterus on magnetic resonance images: Development and performance analysis of an automated deep-learning tool},
  author={Mulliez, Daphn{\'e} and Poncelet, Edouard and Ferret, Laurie and Hoeffel, Christine and Hamet, Blandine and Dang, Lan Anh and Laurent, Nicolas and Ramette, Guillaume},
  journal={Diagnostics},
  volume={13},
  number={16},
  pages={2662},
  year={2023},
  publisher={MDPI}
}

@inproceedings{hatamizadeh2021swin,
  title={Swin unetr: Swin transformers for semantic segmentation of brain tumors in mri images},
  author={Hatamizadeh, Ali and Nath, Vishwesh and Tang, Yucheng and Yang, Dong and Roth, Holger R and Xu, Daguang},
  booktitle={Brainlesion, MICCAI},
  pages={272},
  year={2021},
  organization={Springer}
}

@article{Raffone2024MRILeiomyomasSarcomas,
  title        = {Diagnostic accuracy of MRI in the differential diagnosis between uterine leiomyomas and sarcomas: A systematic review and meta-analysis},
  author       = {Raffone, Antonio and Raimondo, Diego and Neola, Daniele and Travaglino, Antonio and Giorgi, Matteo and Lazzeri, Lucia and De Laurentiis, Francesco and Carravetta, Carlo and Zupi, Errico and Seracchioli, Renato and Casadio, Paolo and Guida, Maurizio},
  journal      = {International Journal of Gynecology \& Obstetrics},
  year         = {2024},
  volume       = {165},
  pages        = {22--33},
  doi          = {10.1002/ijgo.15136}
}

@incollection{Takahama2025DifferentialDiagnosisCervicalCysticLesions,
  author       = {Junko Takahama},
  title        = {Differential Diagnosis of Cervical Cystic Lesions: Nabothian Cyst, Tunnel Cluster, LEGH, MDA},
  booktitle    = {MRI and CT for Decision-Making in Obstetrics and Gynecology Practice},
  editor       = {Noriomi Matsumura and Mitsuru Matsuki and Aki Kido},
  publisher    = {Springer, Singapore},
  year         = {2025},
  pages        = {223--230}
}

@ARTICLE{py06nimg,
  author = {Paul A. Yushkevich and Joseph Piven and Cody Hazlett, Heather and
    Gimpel Smith, Rachel and Sean Ho and James C. Gee and Guido Gerig},
  title = {User-Guided {3D} Active Contour Segmentation of
    Anatomical Structures: Significantly Improved Efficiency and Reliability},
  journal = {Neuroimage},
  year = {2006},
  volume = {31},
  number = {3},
  pages = {1116--1128},
}

@article{paszke2019pytorch,
  title={Pytorch: An imperative style, high-performance deep learning library},
  author={Paszke, Adam and Gross, Sam and Massa, Francisco and Lerer, Adam and Bradbury, James and Chanan, Gregory and Killeen, Trevor and Lin, Zeming and Gimelshein, Natalia and Antiga, Luca and others},
  journal={Advances in neural information processing systems},
  volume={32},
  year={2019}
}

@article{Saleem2025DeepSegmentationUterineMyomas,
  title        = {Deep Learning-Based Automated Segmentation of Uterine Myomas},
  author       = {Saleem, Tausifa Jan and Yaqub, Mohammad},
  journal      = {arXiv},
  volume       = {abs/2508.11010},
  year         = {2025},
  eprint       = {2508.11010},
  archivePrefix= {arXiv},
  primaryClass = {eess.IV},
  doi          = {10.48550/arXiv.2508.11010}
}

@article{Pan2023WAResUNet,
  title        = {WA-ResUNet: A Focused Tail Class MRI Medical Image Segmentation Algorithm},
  author       = {Pan, Haixia and Gao, Bo and Bai, Wenpei and Li, Bin and Li, Yanan and Zhang, Meng and Wang, Hongqiang and Zhao, Xiaoran and Chen, Minghuang and Yin, Cong and Kong, Weiya},
  journal      = {Bioengineering},
  year         = {2023},
  volume       = {10},
  number       = {8},
  pages        = {945},
  doi          = {10.3390/bioengineering10080945}
}

@article{patel2010imaging,
  title={Imaging of endometrial and cervical cancer},
  author={Patel, Shilpa and Liyanage, Sidath H and Sahdev, Anju and Rockall, Andrea G and Reznek, Rodney H},
  journal={Insights into imaging},
  volume={1},
  pages={309--328},
  year={2010},
  publisher={Springer},
  doi={10.1007/s13244-010-0042-7}
}

@ARTICLE{Omi2024-je,
  title     = "Preoperative diagnosis of cervical cystic lesions using magnetic
               resonance imaging: a retrospective study",
  author    = "Omi, Makiko and Tanaka, Yumiko Oishi and Kurihara, Nozomi and
               Sugiyama, Yuko and Tonooka, Akiko and Kanno, Motoko and Fusegi,
               Atsushi and Aoki, Yoichi and Netsu, Sachiho and Abe, Akiko and
               Tanigawa, Terumi and Okamoto, Sanshiro and Nomura, Hidetaka and
               Kanao, Hiroyuki",
  journal   = "BMC Womens. Health",
  publisher = "Springer Science and Business Media LLC",
  volume    =  24,
  number    =  1,
  pages     = "460",
  month     =  aug,
  year      =  2024,
  copyright = "https://creativecommons.org/licenses/by-nc-nd/4.0",
  language  = "en",
  doi={10.1186/s12905-024-03304-8}
}

@ARTICLE{Agostinho2017-ut,
  title     = "{MRI} for adenomyosis: a pictorial review",
  author    = "Agostinho, Lisa and Cruz, Rita and Os{\'o}rio, Filipa and Alves,
               Jo{\~a}o and Set{\'u}bal, Ant{\'o}nio and Guerra, Adalgisa",
  journal   = "Insights Imaging",
  publisher = "Springer Nature",
  volume    =  8,
  number    =  6,
  pages     = "549--556",
  month     =  dec,
  year      =  2017,
  language  = "en",
  doi={10.1007/s13244-017-0576-z}
}

@article{salman2024magnetic,
  title={Magnetic resonance imaging evaluation of gynecological mass lesions: A comprehensive analysis with histopathological correlation},
  author={Salman, Syed and Shireen, Nabeela and Riyaz, Romana and Khan, Sajjad Ahmed and Singh, Janender Pal and Uttam, Anuj},
  journal={Medicine},
  volume={103},
  number={32},
  pages={e39312},
  year={2024},
  publisher={LWW},
  doi={10.1097/MD.0000000000039312}
}

@ARTICLE{Zand2007-yf,
  title     = "Artifacts and pitfalls in {MR} imaging of the pelvis",
  author    = "Zand, Khashayar Rafat and Reinhold, Caroline and Haider, Masoom
               A and Nakai, Asako and Rohoman, Laurian and Maheshwari, Sharad",
  journal   = "J. Magn. Reson. Imaging",
  publisher = "Wiley",
  volume    =  26,
  number    =  3,
  pages     = "480--497",
  month     =  sep,
  year      =  2007,
  copyright = "http://onlinelibrary.wiley.com/termsAndConditions\#vor",
  language  = "en",
  doi={10.1002/jmri.20996}
}

@ARTICLE{Kido2008-zv,
  title     = "Intrauterine devices and uterine peristalsis: evaluation with
               {MRI}",
  author    = "Kido, Aki and Togashi, Kaori and Kataoka, Milliam L and Nakai,
               Asako and Koyama, Takashi and Fujii, Shingo",
  journal   = "Magn. Reson. Imaging",
  publisher = "Elsevier BV",
  volume    =  26,
  number    =  1,
  pages     = "54--58",
  month     =  jan,
  year      =  2008,
  language  = "en",
  doi={10.1016/j.mri.2007.06.001}
}

@article{tong2020recommendations,
  title={Recommendations for MRI technique in the evaluation of pelvic endometriosis: consensus statement from the Society of Abdominal Radiology endometriosis disease-focused panel},
  author={Tong, Angela and VanBuren, Wendaline M and Chami{\'e}, Luciana and Feldman, Myra and Hindman, Nicole and Huang, Chenchan and Jha, Priyanka and Kilcoyne, Aoife and Laifer-Narin, Sherelle and Nicola, Refky and others},
  journal={Abdominal Radiology},
  volume={45},
  pages={1569--1586},
  year={2020},
  publisher={Springer},
  doi={10.1007/s00261-020-02483-w}
}

@article{berczi2025outlier,
  title={Outlier data in volume calculations of uterine fibroids comparing ellipsoid formula and voxel-based segmentation},
  author={B{\'e}rczi, Viktor and Turt{\'o}czki, Kolos Gy{\"o}rgy and Fazekas, Szuzina and Dolla-Tak{\'a}cs, Anna and Stollmayer, R{\'o}bert and Kaposi, P{\'a}l Nov{\'a}k and Kalina, Ildik{\'o} and Budai, Bettina Katalin},
  journal={BMC Medical Imaging},
  volume={25},
  pages={1--8},
  year={2025},
  publisher={Springer},
  doi={10.1186/s12880-025-01672-7}
}

@article{kurban2021uterine,
  title={Uterine artery embolization of uterine leiomyomas: predictive MRI features of volumetric response},
  author={Kurban, Lutfi Ali S and Metwally, Hesham and Abdullah, Mudar and Kerban, Abdurzag and Oulhaj, Abderrahim and Alkoteesh, Jamal Aldeen},
  journal={American Journal of Roentgenology},
  volume={216},
  number={4},
  pages={967--974},
  year={2021},
  publisher={American Roentgen Ray Society},
  doi={10.2214/AJR.20.22906}
}

@article{theis2023deep,
  title={Deep learning enables automated MRI-based estimation of uterine volume also in patients with uterine fibroids undergoing high-intensity focused ultrasound therapy},
  author={Theis, Maike and Tonguc, Tolga and Savchenko, Oleksandr and Nowak, Sebastian and Block, Wolfgang and Recker, Florian and Essler, Markus and Mustea, Alexander and Attenberger, Ulrike and Marinova, Milka and others},
  journal={Insights into Imaging},
  volume={14},
  number={1},
  pages={1},
  year={2023},
  publisher={Springer},
  doi={10.1186/s13244-022-01342-0}
}

@article{pan2024large,
  title={Large-scale uterine myoma MRI dataset covering all FIGO types with pixel-level annotations},
  author={Pan, Haixia and Chen, Minghuang and Bai, Wenpei and Li, Bin and Zhao, Xiaoran and Zhang, Meng and Zhang, Dongdong and Li, Yanan and Wang, Hongqiang and Geng, Haotian and others},
  journal={Scientific Data},
  volume={11},
  number={1},
  pages={410},
  year={2024},
  publisher={Nature Publishing Group UK London},
  doi={10.1038/s41597-024-03170-x}
}

@inproceedings{isensee2024nnu,
  title={nnu-net revisited: A call for rigorous validation in 3d medical image segmentation},
  author={Isensee, Fabian and Wald, Tassilo and Ulrich, Constantin and Baumgartner, Michael and Roy, Saikat and Maier-Hein, Klaus and Jaeger, Paul F},
  booktitle={International Conference on Medical Image Computing and Computer-Assisted Intervention},
  pages={488--498},
  year={2024},
  organization={Springer}, 
  DOI={10.1007/978-3-031-72114-4_47}

}

@article{stewart2017epidemiology,
  title={Epidemiology of uterine fibroids: a systematic review},
  author={Stewart, Elizabeth A and Cookson, CL and Gandolfo, Ruth A and Schulze-Rath, Renate},
  journal={BJOG: An International Journal of Obstetrics \& Gynaecology},
  volume={124},
  number={10},
  pages={1501--1512},
  year={2017},
  publisher={Wiley Online Library},
  DOI={10.1111/1471-0528.14640}
}

@article{khan2014uterine,
  title={Uterine fibroids: current perspectives},
  author={Khan, Aamir T and Shehmar, Manjeet and Gupta, Janesh K},
  journal={International journal of women's health},
  DOI={10.2147/ijwh.s51083},
  pages={95},
  year={2014},
  publisher={Taylor \& Francis}
}

@article{sevindik2023differences,
author = {Sevindik, Betul and Unver Dogan, Nadire and Secilmis, Ozlem and Uysal, Emine and Fazliogullari, Zeliha and Karabulut, Ahmet Kagan},
title = {Differences in the anatomical structure of the uterus between fertile and infertile individuals},
journal = {Clinical Anatomy},
volume = {36},
number = {5},
year = {2023},
pages = {764},
doi = {https://doi.org/10.1002/ca.24045},
}

@article{anneveldt2021lessons,
  title={Lessons learned during implementation of MR-guided High-Intensity Focused Ultrasound treatment of uterine fibroids},
  author={Anneveldt, KJ and Verpalen, IM and Nijholt, IM and Dijkstra, JR and van den Hoed, RD and van’t Veer-ten Kate, M and de Boer, E and van Osch, JAC and Heijman, E and Naber, HR and others},
  journal={Insights into Imaging},
  volume={12},
  pages={1--13},
  year={2021},
  publisher={Springer},
  DOI={10.1186/s13244-021-01128-w}
}

@article{liu2025deep,
  title={Deep learning assisted detection and segmentation of uterine fibroids using multi-orientation magnetic resonance imaging},
  author={Liu, Xin-Yu and Yuan, Zhi-Lin and Cong, Fu-Ze and Mao, Li and Li, Xiu-Li and Zhou, Zhen and Ren, Jing and Li, Yuan and Zhang, Yan and He, Yong-Lan and others},
  journal={Abdominal Radiology},
  pages={1--12},
  year={2025},
  publisher={Springer},
  DOI={10.1007/s00261-025-04934-8}
}

@article{liu2025artificial,
  title={Artificial intelligence for instance segmentation of MRI: advancing efficiency and safety in laparoscopic myomectomy of broad ligament fibroids},
  author={Liu, Feiran and Chen, Minghuang and Pan, Haixia and Li, Bin and Bai, Wenpei},
  journal={Frontiers in Oncology},
  volume={15},
  pages={1549803},
  year={2025},
  DOI={10.3389/fonc.2025.1549803}
}

@article{silva2025automatic,
  title={Automatic flow planning for fetal cardiovascular magnetic resonance imaging},
  author={Silva, Sara Neves and Woodgate, Tomas and McElroy, Sarah and Cleri, Michela and St Clair, Kamilah and Verdera, Jordina Aviles and Payette, Kelly and Uus, Alena and Story, Lisa and Lloyd, David and others},
  journal={Journal of Cardiovascular Magnetic Resonance},
  volume={27},
  number={1},
  pages={101888},
  year={2025},
  publisher={Elsevier},
  DOI={10.1016/j.jocmr.2025.101888}
}

@article{verdera2025heron,
  title={HERON: High-Efficiency Real-Time mOtion quantification and re-acquisitioN for Fetal diffusion MRI},
  author={Verdera, Jordina Aviles and Bortolazzi, Antonia and Silva, Sara Neves and Payette, Kelly and Clair, Kamilah St and McElroy, Sarah and Malik, Shaihan and Hajnal, Joseph and Tomi-Tricot, Raphael and Rutherford, Mary and others},
  journal={IEEE Transactions on Medical Imaging},
  year={2025},
  publisher={IEEE},
  DOI={10.1109/tmi.2025.3569853}
}

@inproceedings{xue2019gadgetron,
  title={Gadgetron inline AI: Effective model inference on MR scanner},
  author={Xue, Hui and Davies, Rhodri and Hansen, David and Tseng, Ethan and Fontana, Marianna and Moon, James C and Kellman, Peter},
  booktitle={Proceedings of the 27th Annual ISMRM Meeting and Exhibition},
  pages={4837},
  year={2019}
}

@inproceedings{chow2021prototyping,
  title={Prototyping image reconstruction and analysis with FIRE},
  author={Chow, K and Kellman, P and Xue, H},
  booktitle={SCMR},
  year={2021}
}

@inproceedings{fraser2011figo,
  title={The FIGO recommendations on terminologies and definitions for normal and abnormal uterine bleeding},
  author={Fraser, Ian S and Critchley, Hilary OD and Broder, Michael and Munro, Malcolm G},
  booktitle={Seminars in reproductive medicine},
  volume={29},
  number={05},
  DOI={10.1055/s-0031-1287662},
  pages={383--390},
  year={2011},
  organization={{\copyright} Thieme Medical Publishers}
}

@INPROCEEDINGS{Khaghani2024-yh,
  title           = "A deep learning-based tool for analyzing the female
                     reproductive system in {MR} images",
  booktitle       = "{ISMRM}",
  author          = "Khaghani, Javad and Basar, Saqib and Chodakiewitz, Yosef
                     and London, Sean and Attariwal, Rajpaul and Hashemi, Sam",
  year            =  2024,
  conference      = "2023 ISMRM",
  location        = "Toronto, ON, Canada"
}

@INPROCEEDINGS{Khaghani2024-pa,
  title           = "An {AI-based} solution for {MR} image analysis of the
                     female reproductive system",
  booktitle       = "{ISMRM}",
  author          = "Khaghani, Javad and Khallaghi, Siavash and Basar, Saqib
                     and Chodakiewitz, Yosef and Attariwala, Rajpaul and
                     Hashemi, Sam",
  year            =  2024,
  address         = "Concord, CA",
  conference      = "2024 ISMRM",
  doi = {https://doi.org/10.1002/ca.24045}
}
\end{document}